\documentclass[11pt]{article}
\usepackage{graphics}
\usepackage{graphicx}
\usepackage{subfloat}
\DeclareGraphicsExtensions{.eps}
\usepackage{float} 
\usepackage{amsmath}
\usepackage{amsfonts}   
\usepackage{amssymb}

\begin{document}
\date{}
\title{{\bf{\Large Magnetic response of holographic Lifshitz superconductors:Vortex and Droplet solutions }}}
\author{
{\bf {{\normalsize Arindam Lala}}$
$\thanks{e-mail: arindam.lala@bose.res.in, arindam.physics1@gmail.com}}\\
 {\normalsize S. N. Bose National Centre for Basic Sciences,}
\\{\normalsize JD Block, Sector III, Salt Lake, Kolkata-700098, India}
\\[0.3cm]}
\maketitle
\begin{abstract}
\small{In this paper a holographic model of $s$-wave superconductor with anisotropic Lifshitz scaling has been considered. In the presence of an external magnetic field our holographic model exhibits both vortex and droplet solutions. Based on analytic methods we have shown that the anisotropy has no effect on the vortex and droplet solutions whereas it may affect the condensation. Our vortex solution closely resembles the Ginzburg-Landau theory and a relation between the upper critical magnetic field and superconducting coherence length has been speculated from this comparison. Using Sturm-Liouville method, the effect of anisotropy on the critical parameters in insulator/superconductor phase transitions has been analyzed.}
\end{abstract}
\section{Introduction and motivations}
The emergence of the AdS/CFT correspondence\cite{Maldacena}-\cite{Witten} has opened up new directions in dealing with the strongly correlated systems. Since its discovery, this duality has been extensively used in several areas in physics such as, fluid/gravity correspondence, QCD, and many others\cite{Policastro}-\cite{Hubney-Minawalla}. In addition, its lucidity and wide range of applicability have led physicists to apply this correspondence in order to understand several strongly coupled phenomena of condensed matter physics\cite{Sachdev-CMT}-\cite{Herzog-CMT}. But in many examples of condensed matter physics it is often observed that the behaviors of the systems are governed by\textit{ Lifshitz-like fixed points}. These fixed points are characterized by the anisotropic scaling symmetry
\begin{equation}
\label{eq:1.1}
t\rightarrow \lambda^{z}t, \qquad x^{i}\rightarrow\lambda x^{i}\qquad   (i=1,2,...,d).
\end{equation}
  The exponent $z$ is called the ``dynamical critical exponent" and it describes the degree of anisotropy between space and time\cite{Chen-LFT}. These are non-Lorentz invariant points and hence the systems are non-relativistic in nature\cite{Brynjolfsson-LHS}.

There have been several attempts to describe these systems holographically using the standard prescriptions of gauge/gravity duality. But due to the nonrelativistic nature of these systems the dual description has been modified and it provides a gravity dual for systems which are realized by nonrelativistic CFTs\cite{Son-NRCFT}-\cite{Taylor-NRCFT}. The gravity dual to Lifshitz fixed points is described by the Lifshitz metric\cite{Kachru-NRCFT}\footnote{We shall set the radius of the AdS space to unity ($L=1$) in our analysis.}:
\begin{equation}
\label{eq:1.2}
ds^{2}=-r^{2z}dt^{2}+\frac{dr^{2}}{r^{2}}+r^{2}dx^{i}dx_{i}
\end{equation}
which respects the scale transformation Eq.\eqref{eq:1.1} along with an additional scaling $r\rightarrow\lambda^{-1}r$. In the limit $z=1$ it gives the $AdS_{d+2}$ metric. On the other hand exact black hole solutions in the asymptotically Lifshitz space-time have been found\cite{Pang-LBH}-\cite{Gonzalez-LBH}.

 Recently, the AdS/CFT duality has been used to understand diverse properties of high $T_{c}$ superconductors \cite{Gubser-HS-1}-\cite{HHH-HS-2}. The studies of these holographic models of superconductor have been extended by including several higher derivative corrections to the usual Einstein gravity as well as in the Maxwell gauge sector (see Ref.\cite{Lala} and references therein). Along with this the response of the holographic superconductors in external magnetic fields has also been studied\cite{Lala}-\cite{Cui-mag}. These studies show interesting vortex and droplet solutions for these models\cite{Albash-mag-3}-\cite{Zeng-mag},\cite{Maeda-mag-2},\cite{Dibakar-mag}. Very recently promising conclusions have been drawn regarding the effects of various corrections to the Einstein-Maxwell sector on the aforementioned solutions\cite{Dibakar-mag-2}-\cite{Dibakar-mag-3}. 

Over the past few years a series of works have been attempted to understand various properties of HS with Lifshitz scaling\cite{Brynjolfsson-LHS},\cite{Cai-LHS}-\cite{Abdalla-LHS}. Very recently the authors of Refs.\cite{Lu-LHS}, \cite{Wu-LHS}  found out interesting effects of anisotropy on the characterizing properties of HS with Lifshitz scaling as well as the effects of external magnetic fields on them. In spite of these attempts several other important issues have been overlooked which we intend to study in the present paper. Thus the motivations of the present analysis may be put forward as follows: $(i)$ It has been confirmed that the anisotropic scaling plays an important role in affecting the behavior of the holographic condensates\cite{Lu-LHS}, \cite{Wu-LHS}. Thus it would be interesting to verify whether the vortex and/or droplet solutions are affected by it; and $(ii)$ It would be nontrivial to study the effects of anisotropy on the holographic condensates. Also, the response of the critical parameters of phase transition to this anisotropy can be studied. 

The paper is organized as follows: In Section 2 we have developed the vortex lattice solution for the $s$-wave superconductors in a Lifshitz black hole background. In Section 3 we have obtained a holographic droplet solution for this superconductor in the Lifshitz soliton background. Finally, in Section 4 we have drawn conclusions and discussed some of the future scopes.
\section{Vortex solution}
In this paper we shall construct a $(2+1)$-dimensional holographic superconductor with Lifshitz scaling (Eq.\eqref{eq:1.1}). According to the gauge/gravity duality the gravitational dual to this model will be a $(3+1)$-dimensional Lifshitz space-time with the following action\cite{Taylor-NRCFT}:
\begin{equation}
\label{eq:2.1}
\mathcal{S}=\frac{1}{16\pi G}\int d^{4}x \sqrt{-g} \Big(R-2\Lambda - \frac{1}{2}\partial_{\mu}\phi\partial^{\mu}\phi-\frac{1}{4}e^{b\phi}\mathcal{F}_{\mu\nu}\mathcal{F}^{\mu\nu}\Big)
\end{equation}
The background over which we intend to work is given by the following four dimensional Lifshitz black hole\cite{Pang-LBH},\cite{Lu-LHS}:
\begin{equation}
\label{eq:2.2}
ds^{2}=-\dfrac{\beta^{2z}}{u^{2z}}f(u)dt^{2}+\dfrac{\beta^{2}}{u^{2}}(dx^{2}+dy^{2})+\dfrac{du^{2}}{u^{2}f(u)}
\end{equation}
where we have chosen a coordinate $u=\frac{1}{r}$, such that the black hole horizon is at $u=1$ and the boundary ($r\rightarrow\infty$) is at $u=0$, for mathematical simplicity. In Eq.\eqref{eq:2.2}
\begin{equation}
\label{eq:2.3}
f(u)=1-u^{z+2},\qquad\qquad \beta(T)=\Big(\dfrac{4\pi T}{z+2}\Big)^{\frac{1}{z}}
\end{equation}
$T$ being the Hawking temperature of the black hole.

The matter action for our model may be written as\cite{HHH-HS-1}\footnote{We are working in the probe limit where gravity and matter decouple and the backreaction of the matter fields (the charged gauge field and the charged massive scalar field) on the background geometry can be neglected. This simplifies the problem without affecting the physical properties of the system. },
\begin{equation}
\label{eq:2.4}
\mathcal{S}_{M}=\int d^{4}x\sqrt{-g}\Big(-\frac{1}{4}F_{\mu\nu}F^{\mu\nu}-|\mathcal{D}_{\mu}\psi|^{2}-m^{2}|\psi|^{2}\Big)
\end{equation}
where $F_{\mu\nu}=\partial_{\mu}A_{\nu}-\partial_{\nu}A_{\mu}$, $\mathcal{D}_{\mu}=\partial_{\mu}-iA_{\mu}$ ($\mu,\nu=t,x,y,u$) and $m$ is the mass of the scalar field $\psi$.

The equations of motion for the scalar field, $\psi$, and the gauge field, $A_{\mu}$, can be obtained from Eq.\eqref{eq:2.4} as
\begin{eqnarray}
\label{eq:2.5}
\dfrac{1}{\sqrt{-g}}\partial_{\mu}\left(\sqrt{-g}\partial^{\mu}\psi\right)-A_{\mu}A^{\mu}\psi-m^{2}\psi-iA^{\mu}\partial_{\mu}\psi-\dfrac{i}{\sqrt{-g}}\partial_{\mu}\left(\sqrt{-g}A^{\mu}\psi\right)=0,\\
\dfrac{1}{\sqrt{-g}}\partial_{\mu}\left(\sqrt{-g}F^{\mu\nu}\right)=j^{\nu}\equiv i\left(\psi^{*}\partial^{\nu}\psi-\psi(\partial^{\nu}\psi)^{*}\right)+2A^{\nu}|\psi|^{2}.
\end{eqnarray}

In order to proceed further we shall consider the following ansatz for the gauge field\cite{Maeda-mag-1}:
\begin{equation}
 \label{eq:2.6}
 A_{\mu}=(A_{t},A_{x},A_{y},0).
 \end{equation} 

We shall make further assumption that the solutions are stationary i.e. independent of time $t$. Using these we may write Eqs.\eqref{eq:2.5},(8) as a set of coupled differential equations given by
\begin{subequations}
\label{eq:2.7}
\begin{align}
\label{eq:2.7.1}
\left(u^{3-z}\partial_{u}\dfrac{f(u)}{u^{z+1}}\partial_{u}+\dfrac{A_{t}^{2}}{\beta^{2z}f(u)}-\dfrac{m^{2}}{u^{2z}}\right)\psi &=\dfrac{-1}{\beta^{2}u^{2z-2}}\Big(\delta^{ij}\mathcal{D}_{i}\mathcal{D}_{j}\psi\Big)\\
\label{eq:2.7.2}
f(u)\beta^{2}\partial_{u}\Big(u^{z-1}(\partial_{u}A_{t})\Big)+u^{z-1}\bigtriangleup A_{t}&=\dfrac{2\beta^{2}A_{t}}{u^{3-z}}\psi^{2}.
\end{align}
\end{subequations}
where $i,j=x,y$ and $\bigtriangleup =\partial_{x}^{2}+\partial_{y}^{2}$ is the \textit{Laplacian operator}.

In order to solve the above set of equations we shall invoke the following boundary conditions\cite{Maeda-mag-1}:

(i) At the asymptotic boundary $(u\rightarrow 0)$, the scalar field $\psi$ behaves as\cite{Lu-LHS}
\begin{equation}
\label{eq:2.8}
\psi\sim C_{1}u^{\triangle_{-}}+C_{2}u^{\triangle_{+}}
\end{equation}
where $\triangle_{\pm}=\frac{(z+2)\pm\sqrt{(z+2)^{2}+4m^{2}}}{2}$ and the coefficients $C_{1}, C_{2}$ are related to the expectation values of the operators dual to $\psi$ with scaling dimension $\triangle_{-}$ and $\triangle_{+}$ respectively. For our analysis we shall always choose the \textit{mass-squared}, $m^{2}$, of the scalar field above its lower bound given by $m_{LB}^{2}=\frac{-(z+2)^{2}}{4}$\cite{Lu-LHS}. With this condition both the modes are normalizable and we may choose either one of them as the expectation value of the dual operator while the other behaves as the source. For the rest of our analysis we shall choose $C_{1}=0$. Also, $\psi$ is regular at the horizon, $u=1$.

(ii) The asymptotic values of the gauge field $A_{\mu}$ give the chemical potential ($\mu$) and the external magnetic field ($\mathcal{B}$) as 
\begin{equation}
\label{eq:2.9}
\mu=A_{t}(\vec{x},u\rightarrow 0),\qquad\qquad\qquad \mathcal{B}=F_{xy}(\vec{x},u\rightarrow 0)
\end{equation}
where $\vec{x}=x,y$.
The regularity of the gauge fields demand that $A_{t}=0$ and  $A_{i}$ is regular everywhere on the horizon.

We shall further regard the external magnetic field as the only tuning parameter of our theory. Following this we shall assume $\mu$ and $T$ of the boundary theory to be fixed and change only $\mathcal{B}$. Considering our model of holographic superconductor analogous to ordinary type-II superconductor, there exists an upper critical magnetic field, $\mathcal{B}_{c_{2}}$, below which the condensation occurs while above the $\mathcal{B}_{c_{2}}$ superconductivity breaks down.

As a next step, define the deviation parameter $\epsilon$ such that\cite{Maeda-mag-1}
\begin{equation}
\label{eq:2.10}
\epsilon=\dfrac{\mathcal{B}_{c_{2}}-\mathcal{B}}{\mathcal{B}_{c_{2}}},\qquad\qquad\epsilon\ll1.
\end{equation}
Let us expand the scalar field $\psi$, the gauge field $A_{\mu}$ and the current $j_{\mu}$ as the following power series in $\epsilon$:
\begin{subequations}\label{eq:2.11}
\begin{align}
\label{eq:2.11:1}
\psi(\vec{x},u)&=\epsilon^{1/2}\psi_{1}(\vec{x},u)+\epsilon^{3/2}\psi_{2}(\vec{x},u)+...,
\\
\label{eq:2.11:2}
A_{\mu}(\vec{x},u)&=A_{\mu}^{(0)}+\epsilon A_{\mu}^{(1)}(\vec{x},u)+...,
\\
\label{eq:2.11.3}
j_{\mu}(\vec{x},u)&=\epsilon j_{\mu}^{(1)}(\vec{x},u)+\epsilon^{2} j_{\mu}^{(2)}(\vec{x},u)+...
\end{align}
\end{subequations}
From Eqs.\eqref{eq:2.10} and \eqref{eq:2.11} we may infer the following interesting points:

(i) Since we have chosen $\epsilon\ll 1$, we are in fact very close to the critical point,

(ii) The positivity of the deviation parameter implies that $\mathcal{B}_{c_{2}}$ is always greater than the applied magnetic field $\mathcal{B}$. This ensures that there is always a non-trivial scalar condensation in the theory that behaves as the order parameter.

Another important point that must be stressed is that, in Eq.\eqref{eq:2.11:2} $A_{\mu}^{(0)}$ is the solution to the Maxwell's equation in the absence of scalar condensate $(\psi=0)$. For the rest of our analysis we shall choose the following ansatz:
\begin{equation}
\label{eq:2.12}
A_{\mu}^{(0)}=\left(A_{t}^{0}(u), 0,A_{y}^{0}(x),0\right).
\end{equation}

Now matching the coefficients of $\epsilon^{0}$ on both sides of  Eq.\eqref{eq:2.7.2} we may obtain,
\begin{equation}
\label{eq:2.13}
A_{t}^{0}=\mu(1-u^{2-z}),\qquad\qquad A_{x}^{0}=0,\qquad\qquad A_{y}^{0}=\mathcal{B}_{c_{2}}x.
\end{equation}

On the other hand using Eqs.\eqref{eq:2.11:1},\eqref{eq:2.13} and using the following ansatz for $\psi_{1}(\vec{x},u)$\cite{Maeda-mag-1}
\begin{equation}
\label{eq:2.14}
\psi_{1}(\vec{x},u)=e^{ipy}\phi(x,u;p)
\end{equation}
where $p$ is a constant, we can write Eq.\eqref{eq:2.7.1} as,
\begin{equation}
\label{eq:2.15}
\left(u^{3-z}\partial_{u}\dfrac{f(u)}{u^{z+1}}\partial_{u}+\dfrac{(A_{t}^{(0)}(u))^{2}}{\beta^{2z}f(u)}-\dfrac{m^{2}}{u^{2z}}\right)\phi(x,u;p) =\dfrac{1}{\beta^{2}u^{2z-2}}\Big[\partial_{x}^{2}+(p-\mathcal{B}_{c_{2}}x)^{2}\Big]\phi(x,u;p).
\end{equation}

We may solve Eq.\eqref{eq:2.15} by using the method of separation of variables\cite{Maeda-mag-1}. In order to do so we shall separate the variable $\phi(x,u;p)$ as follow:
\begin{equation}
\label{eq:2.16}
\phi(x,u;p)=\alpha_{n}(u)\gamma_{n}(x;p)
\end{equation}
with  the separation constant $\lambda_{n}$ ($n=0,1,2,...$).

Substituting Eq.\eqref{eq:2.16} into Eq.\eqref{eq:2.15} we may write the equations for $\alpha_{n}(u)$ and $\gamma(x;p)$ as
\begin{subequations}
\label{eq:2.17}
\begin{align}
\label{eq:2.17.1}
u^{2-2z}f(u)\alpha_{n}^{''}(u)-\left[\dfrac{(z+1)f(u)}{u^{2z-1}}+(z+2)u^{3-z}\right]\alpha_{n}^{'}(u)-\dfrac{m^{2}}{u^{2z}}\alpha(u)+\dfrac{(A_{t}^{(0)})^{2}}{\beta^{2z}f(u)}\alpha(u)&=\dfrac{\lambda_{n}\mathcal{B}_{c_{2}}}{\beta^{2}u^{2z-2}}\alpha_{n}(u),\\
\label{eq:2.17.2}
\left(\partial_{X}^{2}-\dfrac{X^{2}}{4}\right)\gamma_{n}(x;p)&=\dfrac{\lambda_{n}}{2}\gamma_{n}(x;p).
\end{align}
\end{subequations}
where we have identified $X=\sqrt{2\mathcal{B}_{c_{2}}}\left(x-\frac{p}{\mathcal{B}_{c_{2}}}\right)$. Following Ref.\cite{Albash-mag-1} we can write the solutions of Eq.\eqref{eq:2.17.2} in terms of \textit{Hermite functions}, $H_{n}$, with eigenvalue $\lambda_{n}=2n+1$ as
\begin{equation}
\label{eq:2.18}
\gamma_{n}(x;p)=e^{-X^{2}/4}H_{n}(X).
\end{equation}

Note that we have considered $\lambda_{n}$ to be an odd integer. Since the Hermite functions decay exponentially with increasing $X$, which is the natural physical choice, our consideration is well justified\cite{Albash-mag-1}. Moreover, $\lambda_{n}=1$ corresponds to the only physical solution for our analysis. Thus we shall restrict ourselves to the $n=0$ case. With this choice Eq.\eqref{eq:2.18} can be written as
\begin{equation}
\label{eq:2.19}
\gamma_{0}(x;p)=e^{-X^{2}/4}\equiv \text{exp}\left[-\dfrac{\mathcal{B}_{c_{2}}}{2}\left(x-\frac{p}{\mathcal{B}_{c_{2}}}\right)^{2}\right].
\end{equation}

From the above analysis it is clear that $\lambda_{n}$ is independent of the constant $p$. Therefore, a linear combination of the solutions $e^{ipy}\alpha_{0}(u)\gamma_{0}(x;p)$ with different values of $p$ is also a solution to the EoM for $\psi_{1}$. Thus, following this proposition, we obtain
\begin{equation}
\label{eq:2.20}
\psi_{1}(\vec{x},u)=\alpha_{0}(u)\sum_{l=-\infty}^{\infty}c_{l}e^{ip_{l}y}\gamma_{0}(x;p_{l}).
\end{equation}
At this point of discussion it is interesting to note that Eq.\eqref{eq:2.20} is very similar to the expression for the order parameter of the Ginzburg-Landau (G-L) theory of type-II superconductors in the presence of a magnetic field\cite{Tiley}
\begin{equation}
 \label{eq:2.21}
 \psi_{G-L}=\sum_{l}c_{l}e^{ip_{l}y}\text{exp}\left[-\dfrac{(x-x_{l})^{2}}{2\xi^{2}}\right]
 \end{equation} 
where $x_{l}=\frac{k\Phi_{0}}{2\pi\mathcal{B}_{c_{2}}}$, $\Phi_{0}$ being the \textit{flux quanta} and $\xi$ is the \textit{superconducting coherence length}.
Comparing Eq.\eqref{eq:2.21} with Eq.\eqref{eq:2.19} we may obtain the following relation between the critical magnetic field and the coherence length as
\begin{equation}
\label{eq:2.22}
\mathcal{B}_{c_{2}}\propto \dfrac{1}{\xi^{2}}
\end{equation}
which is indeed in good agreement with the result of the G-L theory\cite{Tiley}.

We may obtain the \textit{vortex lattice solution} by appropriately choosing $c_{l}$ and $p_{l}$. In order to do so we shall assume periodicity both in the $x$ and $y$ directions characterized by two arbitrary parameters $a_{1}$ and $a_{2}$. The periodicity in the $y$ direction may be expressed as
\begin{equation}
\label{eq:2.23}
p_{l}=\dfrac{2\pi l}{a_{1}\xi},\qquad\qquad\qquad l\in \mathbb{Z}.
\end{equation}

Using Eqs.\eqref{eq:2.22},\eqref{eq:2.23} we may rewrite Eq.\eqref{eq:2.19} for different values of $l$ as
\begin{equation}
\label{eq:2.24}
\gamma(x,y)=\sum_{l=-\infty}^{\infty}c_{l}\;\text{exp}\left(\dfrac{2\pi i l y}{a_{1}\xi}\right)\text{exp}\left[-\dfrac{1}{2\xi^{2}}\left(x-\frac{2\pi l \xi}{a_{1}}\right)^{2}\right]
\end{equation}
where the coefficient $c_{l}$ may be chosen as
\begin{equation}
\label{eq:2.25}
c_{l}=\text{exp}\left(\dfrac{-i\pi a_{2}l^{2}}{a_{1}^{2}}\right).
\end{equation}

As a next step, we rewrite Eq.\eqref{eq:2.24} by using the \textit{elliptic theta function}, $\vartheta_{3}(v,\tau)$,\footnote{$\vartheta(v,\tau)=\sum_{l=-\infty}^{\infty}\text{exp}\left(2i\pi v l+i\pi \tau l^{2}\right)$} as
\begin{equation}
\label{eq:2.26}
\psi_{1}(\vec{x},u)=\alpha_{0}(u)\text{exp}\left(\dfrac{-x^{2}}{2\xi^{2}}\right)\vartheta_{3}(v,\tau).
\end{equation}
where $v$ and $\tau$ may be identified as
\begin{equation}
\label{eq:2.27}
v=\dfrac{y-ix}{a_{1}\xi},\qquad\qquad\qquad \tau=\dfrac{2\pi i-a_{2}}{a_{1}^{2}}.
\end{equation}

Following Refs.\cite{Maeda-mag-2},\cite{Dibakar-mag-3} and using the pseudo-periodicity of $\vartheta_{3}(v,\tau)$ we see that the function $\sigma(\vec{x})\equiv |\text{exp}\left(\dfrac{-x^{2}}{2\xi^{2}}\right)\vartheta_{3}(v,\tau)|^{2}$ represents a vortex lattice in which the fundamental region is spanned by the following two lattice vectors
\begin{equation}
\label{eq:2.28}
\vec{v}_{1}=a_{1}\xi \partial_{y},\qquad\qquad\qquad \vec{v}_{2}=\dfrac{2\pi \xi}{a_{1}}\partial_{x}+\dfrac{a_{2}}{a_{1}}\partial_{y}.
\end{equation}

We may put forward the main results of this section as follows:

(i) From Eq.\eqref{eq:2.26} it is observed that the vortex solution does not depend upon the dynamic exponent $z$. This suggests that, whether the boundary field theory is relativistic or non-relativistic, the vortex structure remains the same. Although it is interesting to note that the exponent $z$ may have non-trivial effects on the condensation of the scalar field as is evident from Eq.\eqref{eq:2.17.1}.

(ii) Eq.\eqref{eq:2.26} also suggests that the structure of the vortex lattice is indeed controlled by the superconducting coherence length, $\xi$. Moreover, the solution has a \textit{Gaussian profile} along the $x$ direction. As the coherence length decreases the lattice structure gradually dies out. This behavior is similar to that of ordinary type-II superconductors\cite{Tiley}.
\section{Droplet solution}
In this section we shall consider the holographic insulator/superconductor phase transition in the asymptotically Lifshitz space-time\footnote{The insulator/superconductor phase transition is realized in the CFT language as a phase transition in which a large enough $U(1)$ chemical potential, $\mu$, overcomes the mass gap related to the scalar field, $\psi$. This mechanism allows $\psi$ to condensate above a critical value, $\mu_{c}$. In fact a soliton background, which includes an extra compactified spatial direction, precisely generates this mass gap resembling an insulating phase. }. Our primary goal will be to extract the droplet solution for the $s$-wave holographic Lifshitz superconductor. In order to perform our analysis we shall consider a planar \textit{Lifshitz soliton}\footnote{The Lifshitz soliton solution is obtained by performing a double Wick rotation of the Lifshitz black hole solution\cite{Lu-LHS},\cite{Way}.} background in $5$-dimensions which can be written as\cite{Lu-LHS},\cite{Way},
\begin{equation}
\label{eq:2.29}
ds^{2}=-r^{2}dt^{2}+r^{2}(dx^{2}+dy^{2})+\dfrac{dr^{2}}{r^{2}f(r)}+r^{2z}f(r)d\chi^{2}
\end{equation}
where $f(r)=\left(1-\frac{1}{r^{z+3}}\right)$ and the spatial direction $\chi$ is compactified to a circle and has a periodicity $\chi = \chi +\pi$. The geometry looks like a cigar in the $(r,\chi)$ directions. However, it will be more convenient to work in polar coordinates, $x=\rho sin\theta,\;y=\rho cos\theta$\cite{Cai-mag},\cite{Dibakar-mag}. With this choice Eq.\eqref{eq:2.29} can be written as,
\begin{equation}
\label{eq:2.30}
ds^{2}=-r^{2}dt^{2}+r^{2}(d\rho^{2}+\rho^{2}d\theta^{2})+\dfrac{dr^{2}}{r^{2}f(r)}+r^{2z}f(r)d\chi^{2}.
\end{equation}

We shall consider Maxwell-scalar action in $5$ dimensions as the matter action of our theory which is written as\cite{HHH-HS-1}\footnote{We shall again work in the probe limit.},
\begin{equation}
\label{eq:2.31}
\mathcal{S}_{M}=\int d^{5}x\sqrt{-g}\Big(-\frac{1}{4}F_{\mu\nu}F^{\mu\nu}-|\mathcal{D}_{\mu}\psi|^{2}-m^{2}|\psi|^{2}\Big).
\end{equation}

 In the probe limit we shall choose the following ansatz for the gauge field close to the critical point of phase transition $(\mu\sim\mu_{c},\;\psi\sim 0)$
 \begin{equation}
 \label{eq:2.32}
 A=\mu_{c} dt+\dfrac{1}{2}\mathcal{B}\rho^{2}d\theta
 \end{equation}
where $\mu$ is the chemical potential and $\mathcal{B}$ is the constant external magnetic field related to the vector potential.

The equation of motion for the scalar field, $\psi$, can be derived by varying the action \eqref{eq:2.31} w.r.t. $\psi$ and is given by
\begin{eqnarray}
\label{eq:2.33}
\partial_{r}^{2}F(t,r)+\left(\dfrac{f'(r)}{f(r)}+\dfrac{(z+4)}{r}\right)\partial_{r}F(t,r)-\dfrac{\partial_{t}^{2}F(t,r)}{r^{4}f(r)}+\dfrac{2 i\mu_{c}}{r^{4}f(r)}\partial_{t}F(t,r)+\Bigg[\dfrac{\partial_{\chi}^{2}H(\chi)}{r^{2 z+2}f^{2}(r)H(\chi)}\\ \nonumber
-\dfrac{m^{2}}{r^{2}f(r)}-\dfrac{\mathcal{B}^{2}\rho^{2}}{4r^{4}f(r)}+\dfrac{\mu_{c}^{2}}{r^{4}f(r)}+\dfrac{\partial_{\rho}\Big(\rho\partial_{\rho}U(\rho)\Big)}{r^{4}f(r)U(\rho)\rho}\Bigg]F(t,r)=0
\end{eqnarray}
where we have used Eq.\eqref{eq:2.32} and considered the following ansatz
\begin{equation}
\label{eq:2.34}
\psi(t,r,\chi,\rho)=F(t,r)H(\chi)U(\rho).
\end{equation}
 	
Now, applying the method of separation of variables we finally obtain the following three equations:
\begin{subequations}
\label{eq:2.35}
\begin{align}
\label{eq:2.35.1}
\dfrac{1}{\rho}\partial_{\rho}\left(\rho\partial_{\rho}U(\rho)\right)-\dfrac{1}{4}\mathcal{B}^{2}\rho^{2}U(\rho)&=-k^{2}U(\rho),\\
\label{eq:2.35.2}
\partial_{\chi}^{2}H(\chi)&=-\lambda^{2}H(\chi),\\
\label{eq:2.35.3}
\partial_{r}^{2}F(t,r)+\left(\dfrac{f'(r)}{f(r)}+\dfrac{(z+4)}{r}\right)\partial_{r}F(t,r)-\dfrac{\partial_{t}^{2}F(t,r)}{r^{4}f(r)}+\dfrac{2 i\mu_{c}}{r^{4}f(r)}\partial_{t}F(t,r)\\ \nonumber
+\dfrac{1}{r^{4}f(r)}\Bigg[\mu_{c}^{2}-m^{2}r^{2}-k^{2}-\dfrac{\lambda^{2}}{f(r)r^{2z-2}}\Bigg]F(t,r)&=0,
\end{align}
\end{subequations}
where $\lambda$ and $k$ are some arbitrary constants. 

Eq.\eqref{eq:2.35.2} has the solution of the form
\begin{equation}
\label{eq:2.36}
H(\chi)=\text{exp}(i\lambda \chi)
\end{equation}
which gives $\lambda=2n,\;n\in\mathbb{Z}$, owing to the periodicity of $H(\chi)$ mentioned earlier.

Eq.\eqref{eq:2.35.1} is similar to the equation of a \textit{harmonic oscillator} with $k^{2}=l|\mathcal{B}|$, $l\in\mathbb{Z}^{+}$. We shall expect that the lowest mode of excitation ($n=0,\;l=1$) will be the first to condensate and will give the most stable solution after condensation\cite{Cai-mag}-\cite{Dibakar-mag}. 

At this point let us discuss one of the main results of this paper. From Eq.\eqref{eq:2.35.1} we observe that it has the following solution
\begin{equation}
\label{eq:2.37}
U(\rho)=\text{exp}\left(\dfrac{-|\mathcal{B}|\rho^{2}}{4}\right).
\end{equation}

This suggests that for any finite magnetic field, the holographic condensate will be confined to a finite circular region. Moreover, if we increase the magnetic field this region shrinks to its size and for a large value of the magnetic field this essentially becomes a point at the origin with a nonzero condensate. This is precisely the holographic realization of a superconducting droplet.

As a next step, we shall be interested in solving Eq.\eqref{eq:2.35.3} in order to determine a relation between the critical parameters $(\mu_{c}\; \text{and}\; \mathcal{B})$ in this insulator/superconductor phase transition. In order to do so, we shall further define $F(t,r)=e^{-i\omega t}R(r)$. With this definition we may rewrite  Eq.\eqref{eq:2.35.3} as,
\begin{equation}
\label{eq:2.38}
R''(u)+\left(\dfrac{f'(u)}{f(u)}-\dfrac{z+2}{u}\right)R'(u)+\dfrac{1}{f(u)}\left(\mu_{c}^{2}-\mathcal{B}-\dfrac{m^{2}}{u^{2}}\right)R(u)=0
\end{equation}
where $u=\frac{1}{r}$ and we have put $\omega=0$ since we are interested in perturbations which are marginally stable\cite{Dibakar-mag}. Here `prime' denotes derivative w.r.t. $u$.

We shall choose a trial function $\Lambda(u)$ such that
\begin{equation}
\label{eq:2.39}
R(u\rightarrow 0)\sim \langle\mathcal{O}_{{\triangle}_{+}}\rangle u^{\triangle_{+}}\Lambda(u)
\end{equation}
where  $\triangle_{\pm}=\frac{(z+3)\pm\sqrt{(z+3)^{2}+4m^{2}}}{2}$, $m_{LB}^{2}=\frac{-(z+3)^{2}}{4}$ \cite{Lu-LHS} and $\Lambda(0)=1,\;\Lambda'(0)=0$. Note that we have identified $C_{2}$ in Eq.\eqref{eq:2.8} as the expectation value of the condensation operator, $ \langle\mathcal{O}_{{\triangle}_{+}}\rangle$.

Substituting Eq.\eqref{eq:2.39} into Eq.\eqref{eq:2.38} we finally get,
\begin{equation}
\label{eq:2.40}
\left(\mathcal{P}(u)\Lambda'(u)\right)'+\mathcal{Q}(u)\Lambda'(u)+\Gamma\mathcal{R}(u)\Lambda(u)=0
\end{equation}
where $\Gamma=(\mu_{c}^{2}-\mathcal{B})$ and
\begin{subequations}
\label{eq:2.41}
\begin{align}
\label{eq:2.41.1}
\mathcal{P}&=\left(1-u^{z+3}\right)u^{2\triangle_{+}-z-2}\\
\label{eq:2.41.2}
\mathcal{Q}&=\Big[\triangle_{+}\left(\triangle_{+}-1\right)\left(1-u^{z+3}\right)-m^{2}-\triangle_{+}\left(z+2+u^{z+3}\right)\Big]u^{2\triangle_{+}-z-4}\\
\label{eq:2.41.3}
\mathcal{R}&=u^{2\bigtriangleup_{+}-z-2}.
\end{align}
\end{subequations}

Note that, Eq.\eqref{eq:2.40} is indeed a standard \textit{Sturm-Liouville} eigenvalue equation. Thus, we may write the eigenvalue, $\Gamma$, by using the following formula\cite{Siopsis-SL}
\begin{equation}
\label{eq:2.42}
\Gamma=\dfrac{\int_{0}^{1}du\Big(\mathcal{P}(u)(\Lambda'(u))^{2}+\mathcal{Q}(u)\Lambda^{2}(u)\Big)}{\int_{0}^{1}du\mathcal{P}(u)\Lambda^{2}(u)}=\Gamma(\alpha,z,m^{2})
\end{equation}

where we have chosen $\Lambda(u)=1-\alpha u^{\triangle_{+}}$. Thus we may argue that, unlike the case of usual holographic superconductors\cite{Dibakar-mag}, the quantity $\Gamma=\mu_{c}^{2}-\mathcal{B}$ depends on the dynamic critical exponent ($z$). Therefore we may conclude that the anisotropic scaling indeed manipulates the relation between the parameters of the phase transition. In the Tables \ref{Table:1},\ref{Table:2},\ref{Table:3} below we have shown the non-trivial dependence of $\Gamma$ on $z$.

\begin{table}[h]
\centering
\begin{tabular}{|c||c|c|c|c|c|c|}
\hline 
$m^{2}$ & -3.0 & -2.0 & -1.0 & 1.0 & 2.0 & 3 \\ 
\hline 
$\Gamma$ & 2.41947 & 5.51156 & 7.59806 & 11.1513 & 12.7798 & 14.3487 \\ 
\hline 
\end{tabular} 
\caption{{\label{Table:1}}Variation of $\Gamma$ for $z=\frac{1}{2}$ ($m_{LB}^{2}=-3.0625$)}
\end{table}

\begin{table}[h]
\centering
\begin{tabular}{|c||c|c|c|c|c|c|c|c|}
\hline 
$m^{2}$ & -4.5 & -3.5 & -2.5 & -1.5 & 1.5 & 2.5 & 3.5 & 4.5\\ 
\hline 
$\Gamma$ & 4.96028 & 7.50645 & 9.59179 & 11.479 &   16.5666 & 18.1496 & 19.6951 & 21.2097\\ 
\hline 
\end{tabular} 
\caption{{\label{Table:2}}Variation of $\Gamma$ for $z=\frac{3}{2}$ ($m_{LB}^{2}=-5.06$)}
\end{table}

\begin{table}[h]
\centering
\begin{tabular}{|c||c|c|c|c|c|c|c|c|}
\hline 
$m^{2}$ & -7.0 & -5.0 & -3.0 & -1.0 & 1.0 & 3.0 & 5.0 & 7.0\\ 
\hline 
$\Gamma$ & 5.61026 & 10.5782 & 14.4483 & 17.9369 & 21.2117 & 24.3448 & 27.3750 & 30.3261\\ 
\hline 
\end{tabular} 
\caption{{\label{Table:3}}Variation of $\Gamma$ for $z=\frac{5}{2}$ ($m_{LB}^{2}=-7.5625$)}
\end{table}

\section{Conclusions and future scopes}
In this paper we have focused our attention to the study of a holographic model of $s$-wave superconductor with Lifshitz scaling in the presence of external magnetic field by using the gauge/gravity duality. Working in the probe limit we have constructed vortex and droplet solutions for our holographic model by considering a Lifshitz black hole and a Lifshitz soliton background, respectively. Unlike the AdS/CFT holographic superconductors there is a non-trivial dynamic exponent in the theory which is responsible for an anisotropy between the temporal and the spatial dimensions of the space-time resulting certain noticeable changes of the properties of the superconductor\cite{Brynjolfsson-LHS},\cite{Cai-LHS}-\cite{Wu-LHS}. Also, due to the non-relativistic nature of the filed theory, the model is governed by the AdS/NRCFT correspondence\cite{Son-NRCFT}-\cite{Taylor-NRCFT}.

The primary motivation of the present study is to verify the possibility of vortex and droplet solutions, which are common to the usual holographic superconductors described by the AdS/CFT correspondence\cite{Albash-mag-3}-\cite{Zeng-mag},\cite{Maeda-mag-2},\cite{Dibakar-mag}, for this class of holographic superconductors as well as to consider the effects of anisotropy on these solutions. Based on purely analytic methods we have been able to construct these solutions. Our analysis shows that, although, the anisotropy has no effects on the vortex lattice solutions, it may have a non-trivial effect on the formation of holographic condensates. Also, a close comparison between our results and those of the Ginzburg-Landau theory reveals the fact that the upper critical magnetic field ($\mathcal{B}_{c_{2}}$) is inversely proportional to the square of the superconducting coherence length ($\xi$). This allows us to speculate the behavior of $\mathcal{B}_{c_{2}}$ with temperature although this requires further investigations which is expected to be explored in the future. On the other hand, based on the method of separation of variables, we have been able to model a holographic droplet solution by working in a Lifshitz soliton background and considering insulator/superconductor phase transition. Our analysis reveals that a holographic droplet is indeed formed in the $\rho-\theta$ plane with a non-vanishing condensate. Also, this droplet grows in size until it captures the entire plane when the external magnetic field $\mathcal{B}\rightarrow 0$. Interestingly, it is observed that the anisotropy does not affect the droplet solution. On top of that, we have determined a relation between the critical parameters of the phase transition by using the Sturm-Liouville method\cite{Siopsis-SL}. Interestingly, this relation is solely controlled by the dynamic exponent ($z$) which in turn exhibits the effects of anisotropy on the condensate (cf. Eq.\eqref{eq:2.42}).

Although we have performed detail analytic calculations regarding some subtle issues of holographic Lifshitz superconductors, there might have been even more interesting outcomes that need further explorations. Some of these may be listed as follows:

$(i)$ It will be interesting to carry out an analysis to see whether $p$-wave as well as $d$-wave holographic Lifshitz superconductors form vortex and/or droplet solutions. In this regard the effects of anisotropy on these solutions may be studied. 

$(ii)$ It is observed that the holographic model of superconductor analyzed in this paper is quite similar to the real world high-$T_{c}$ superconductors. But, this is a phenomenological model where we have chosen the fields and their interactions by hand\cite{HHH-HS-2}. We have not provided any microscopic theory which drives our model of superconductivity. It is expected to have a microscopic theory by proper embedding of the model into the string theory.

$(iii)$ Note that there are nontrivial dependencies of the equations of motion on the dynamic exponent $z$ derived from the actions of our model. This encourages us to obtain the free energy and the $R$-current for our model and study the effect(s) of anisotropy on them. More specifically, it will be important to look for any corrections to the \textit{usual Ginzburg-Landau current} due to the presence of anisotropy. In this regard we may also study the long-wavelength limit of the results thus obtained.

 Apart from the points mentioned above there are several other non-trivial issues, such as the study of the effects of dynamical magnetic fields as well as of backreaction\footnote{The author thanks Hong-Qiang Leng for bringing this to attention.}\cite{Kanno-bkr},\cite{Leng-bkr} on the properties of the $s$-wave Lifshitz superconductor, the effects of various non-linear corrections in the gauge and/or gravity sector on this superconductor and the study of the holographic model considered here in higher dimensions, that we wish to illuminate in the future.

\section*{Acknowledgments}
The author would like to thank C.S.I.R, India for financial support (File No.$09/575(0086)/2010-\text{EMR-I}$). He would also like to thank Rabin Banerjee and Dibakar Roychowdhury for useful discussions.

\end{document}